\documentclass{article}
\usepackage{latexsym}
\usepackage{graphics}
\newtheorem{proposition}{Proposition}
\newtheorem{theorem}{Theorem}

\begin{document}
\title{A note on boundary value problems for black hole evolution}
\author{Mihalis Dafermos and Igor Rodnianski}
\maketitle
\begin{abstract}
In recent work of Allen et.~al., heuristic and numerical arguments
were put forth to suggest that boundary 
value problems for black hole evolution, where an appropriate 
Sommerfeld radiation condition is imposed, 
would probably fail to produce Price law tails. The interest in
this issue lies in its possible
implications for numerical relativity, where black hole
evolution is typically studied in terms of such boundary formulations.
In this note, it is shown rigorously that indeed, Price law tails
do not arise in this case, i.e.~that Sommerfeld (and more general) radiation 
conditions lead to decay faster than any polynomial power. Our setting is 
the collapse of a spherically symmetric self-gravitating scalar field.
We allow an additional gravitationally coupled Maxwell field.
The proof also applies to the easier problem of a spherically 
symmetric solution of the wave equation on a Schwarzschild 
or Reissner-Nordstr\"om background.
The method relies on previous work of the authors.
\end{abstract}

One of the characteristic features of black holes forming in the context
of gravitational collapse is the appearance of
so-called \emph{Price law tails}. 
These describe the decay of radiation on timelike curves 
in the exterior region of the black hole, as well as the decay 
of the radiation flux along the event horizon. In the context of
radiation described by a scalar field, Price~\cite{rpr:ns} put forth
heuristic arguments indicating that this decay should be $\sim v^{-3}$, where 
$v$ is an Eddington-Finkelstein-like advanced time coordinate.
In~\cite{mi:mazi}, 
it was rigorously proven in the context of the coupled Einstein-scalar
field equations under spherical symmetry (and more generally
of the Einstein-Maxwell-scalar field equations) that
\begin{equation}
\label{todikomas}
|\phi|+|\partial_v\phi|\le C_{\epsilon,R} v^{-3+\epsilon},
\end{equation}
in the region $r\le R$, for any $\epsilon>0$ and sufficiently large $R$. 
The decay rates $(\ref{todikomas})$ also hold for 
spherically symmetric solutions of the
wave equation on a fixed Schwarzschild or Reissner-Nordstr\"om background.

In the present paper, it is proven that if appropriate boundary 
conditions are imposed at $r=R$, for fixed $R$, then 
decay can be proven at rates faster than any given polynomial power. 
The conditions include standard Sommerfeld boundary conditions.
This paper was motivated by~\cite{abbp:rt}, where the absense
of Price law tails was suggested on the basis of heuristic
arguments and numerical calculations. In particular,
this paper confirms the validity of the heuristics of~\cite{abbp:rt}.

In fact, the argument of this paper is a rather straightforward application
of the method of~\cite{mi:mazi}. In that paper, decay rates were
achieved by an inductive argument, at each step of which, the rate
was improved  by an amount constrained by information 
coming from infinity. As
we shall see below, with appropriate boundary conditions,
there is no such constraint, and the 
decay rate can be improved at each step by a fixed
finite polynomial power.
Thus, the inductive argument yields that any polynomial power can be
achieved. 

\section{The equations}
The arguments of this paper apply to the Einstein-Maxwell-scalar field
system under spherical symmetry. 
For an introduction to this system, the author can 
consult~\cite{md:si, md:cbh}. 
The Einstein-scalar field system, whose rigorous study
was initiated in~\cite{chr:sgsf}, is of course a special case, where
the Maxwell part vanishes. 
The results of this paper also apply in the case where 
the wave equation is decoupled from the Einstein equations. For
more on the relationship between these two problems, see~\cite{mi:mazi}.

Recall that the quotient manifold $\mathcal{Q}=\mathcal{M}/SO(3)$
inherits a $1+1$-dimensional Lorentzian metric, which in null coordinates
takes the form $-{\Omega^2}du dv$. The metric of $\mathcal{M}$
can then be written $-\Omega^2dudv+r^2\gamma$, 
where $\gamma$ is the standard metric on $S^2$, and $r$ is
the area radius function $r:\mathcal{Q}\to{\bf R}$ defined
by $r(q)=\sqrt{Area(q)/4\pi}$. We recall the Hawking mass
$m=\frac r2\left(1+4\Omega^{-2}\partial_ur\partial_vr\right)$, and the
mass aspect function $\mu=\frac{2m}r$. Under spherical symmetry,
the Maxwell equations decouple, contributing a 
$\frac{e^2}{r^4}\Omega^2dudv+\frac{e^2}{2r^2}\gamma$
term to the energy momentum tensor, where $e$ is a constant.
Defining the ``renormalized'' Hawking mass $\varpi=m+\frac{e^2}{2r}$,
the evolution of the metric and scalar field are then completely
determined by the following sytem of equations:
\begin{equation}
\label{ruqu}
\partial_u{r}=\nu,
\end{equation}
\begin{equation}
\label{rvqu}
\partial_v{r}=\lambda,
\end{equation}
\begin{equation}
\label{lqu}
\partial_u\lambda=\lambda\left(\frac{2\nu}{1-\mu}\left(\frac{\varpi}{r^2}
-\frac{e^2}{r^3}\right)
\right),
\end{equation}
\begin{equation}
\label{nqu}
\partial_v\nu=\nu\left(\frac{2\lambda}{1-\mu}\left(\frac{\varpi}{r^2}
-\frac{e^2}{r^3}\right)
\right),
\end{equation}
\begin{equation}
\label{puqu}
\partial_u \varpi=\frac{1}{2}(1-\mu)\left(\frac\zeta\nu\right)^2\nu,
\end{equation}
\begin{equation}
\label{pvqu}
\partial_v \varpi=\frac{1}{2}(1-\mu)\left(\frac\theta\lambda\right)^2\lambda,
\end{equation}
\begin{equation}
\label{sign1}
\partial_u\theta=-\frac{\zeta\lambda}r,
\end{equation}
\begin{equation}
\label{sign2}
\partial_v\zeta=-\frac{\theta\nu}r.
\end{equation}

We also easily obtain from the above the equations
\begin{equation}
\label{fdb1}
\partial_u\frac{\lambda}{1-\mu}=
\frac1{r}\left(\frac\zeta\nu\right)^2\nu\frac{\lambda}{1-\mu}
\end{equation}
and
\begin{equation}
\label{fdb2}
\partial_v\frac{\nu}{1-\mu}=
\frac{1}r\left(\frac\theta\lambda\right)^2\lambda\frac{\nu}{1-\mu}.
\end{equation}

\section{The assumptions}
We will assume the existence of a spacetime bounded by an ingoing null curve,
a timelike curve, and an outgoing null curve, which we will call the
\emph{event horizon}.\footnote{Compare with~\cite{abbp:rt},
where an additional inner boundary
condition was imposed on a timelike curve ``near'' the event
horizon. This was necessitated by the fact that the Regge-Wheeler coordinates
used in~\cite{abbp:rt} put the event horizon $\mathcal{H}$ at $r_*=-\infty$,
even though $\mathcal{H}$ is part of the spacetime.} 
Certain conditions will be imposed on the initial 
ingoing null curve and the timelike boundary. 
In this paper, we will not concern ourselves with the explicit
\emph{construction} of solutions satisfying the conditions described here, 
although this can be accomplished easily enough 
using the methods of~\cite{md:sssts, mi:mazi}.

We consider a spacetime $\mathcal{Q}$ 
with Penrose diagram depicted below,
where $\Gamma$ is a timelike boundary:
\[
\includegraphics{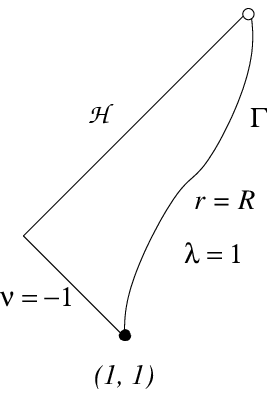}
\]
Here, $\Gamma\cup\mathcal{H}\subset\mathcal{Q}$, but
$\Gamma\cap\mathcal{H}=\emptyset$.
The point of intersection
of the initial ingoing ray and $\Gamma$ will be $(1,1)$. The ingoing
ray will then be the curve $v=1$.  
The equations $(\ref{ruqu})$--$(\ref{sign2})$ are assumed to hold
pointwise. 
On $\Gamma$, $\lambda=1$, $r=R$ for some constant $R$, 
and $m$ will be non-increasing as a function of $v$,
such that $M_0=2\sup_{\Gamma} m<R$. Moreover, we assume that $v$ ranges
in all of $[1,\infty)$.
On the ingoing null segment $v=1$, we assume
$\nu=-1$, $\zeta$ bounded,
$r>c>0$. 
This makes $\{v=1\}\cap\mathcal{H}=
(\tilde{U},1)$ for some $\tilde{U}<\infty$. We will assume $m\ge0$
at $(\tilde{U},1)$.

We first show
\begin{proposition}
Under the above assumptions, $\nu<0$ in $\mathcal{Q}$, 
and $\lambda>0$ in $\mathcal{Q}\setminus\mathcal{H}$.
\end{proposition}

\noindent\emph{Proof.} Let $\mathcal{U}$ be the region
\[
\mathcal{U}=\{p:\nu<0{\rm\ in\ } \overline{J^{-}(p)}\cap\mathcal{Q}\}
\]
We first note that $\nu$ is continuous. For $\nu$ is assumed
to be differentiable in the $v$ direction, while differentiating
$(\ref{nqu})$ in the $u$ direction one obtains an evolution equation
for $\partial_u\nu$, which initially is $0$. 
By continuity of $\nu$, $\mathcal{U}$ is an open subset of $\mathcal{Q}$.
In $\overline\mathcal{U}\cap\mathcal{Q}$, we clearly have $r\le R$,
and $r<R$ in $\overline\mathcal{U}\cap\mathcal{Q}\setminus\Gamma$.
Thus, it follows that
$\lambda>0$ in $\mathcal{U}\setminus\mathcal{H}$. 
For, if $\lambda(u',v')\le0$, and
$r(u',v')<R$, then $r(u',v^*)<R$ for all $v^*>v'$, and thus
the curve $u=u'$ cannot intersect $\Gamma$.\footnote{This statement
follows from the Raychaudhuri equation $\partial_v(\Omega^{-2}\partial_vr)
=-r\Omega^{-2}T_{vv}\le0$. See~\cite{md:sssts}}
Thus, by $(\ref{pvqu})$, 
we have $\partial_v\varpi\ge0$ in $\overline{\mathcal{U}}$,
and so, denoting 
$\Pi_1=\inf_{u\in[1,\tilde{U}]}\varpi(u,1)$
we have $\Pi_1\le\varpi$ in $\overline{\mathcal{U}}\cap\mathcal{Q}$.
Integrating $\partial_vr=\lambda$ from $v=1$,
it is clear that $r\ge c$. From $(\ref{fdb1})$, we have 
\begin{equation}
\label{alloasteri}
\frac{\lambda}{1-\mu}\le \frac{1}{1-\frac{M_0}R}
\end{equation}
Integrating $(\ref{nqu})$, it follows that
\[
-\nu(u,v)\ge\tilde{C}(c,M_0,R,e,\Pi,v)>0
\]
on $\overline{\mathcal{U}}\cap\mathcal{Q}$. 
Thus $\mathcal{U}=\overline\mathcal{U}\cap\mathcal{Q}=\mathcal{Q}$.
This completes the proof. $\Box$

We note now that from the above Proposition and the
results of~\cite{md:sssts}, it follows that $\partial_vm\ge0$, 
$\partial_um\le0$ in $\mathcal{Q}$, and thus, since
$m(\tilde{U},1)\ge0$, we have $m\ge0$ throughout $\mathcal{Q}$.  This
immediately yields
\begin{equation}
\label{asteri}
0\le1-\mu\le1,
\end{equation}
where we use the fact that $\lambda$ and $1-\mu$ have the same
sign to obtain the lower bound.

\section{The decay theorem}
We will show the following
\begin{theorem}
For any $p>0$, there exists a constant $C(p)$ such that
\[
|\theta(u,v)|\le Cv^{-p}.
\]
\end{theorem}
Consider the statement
\begin{enumerate}
\item[$(S_q)$]
There exists a constant $C(q)$ such that
\[
|\theta(u,v)|\le Cv^{-q}
\]
\begin{equation}
\label{evergeia}
\int_v^\infty{\theta^2(\tilde{U},\bar{v})d\bar{v}}\le Cv^{-2q}.
\end{equation}
\end{enumerate}
We shall prove the following two propositions:
\begin{proposition}
$S_0$ is true.
\end{proposition}
\begin{proposition}
$S_q$ implies $S_{q+\frac14}$.
\end{proposition}
The above two propositions and the Archimedean property of the real
numbers immediately yield the Theorem. $\Box$
\vskip.5pc
\noindent\emph{Proof of Proposition 2}. We note that Proposition 1
gives all the information necessary to apply Section 5
of~\cite{mi:mazi}. The propositions of that section
immediately yield the result of this proposition.
In the process, one obtains a lower bound
\begin{equation}
\label{katwfragma}
\frac{\lambda}{1-\mu}\ge c'>0,
\end{equation}
which we will require below.
$\Box$
\vskip.5pc
\noindent\emph{Proof of Proposition 3}. Consider a dyadic decomposition
of the event horizon, with respect to the $v$ coordinate, as in~\cite{mi:mazi}.
\[
\includegraphics{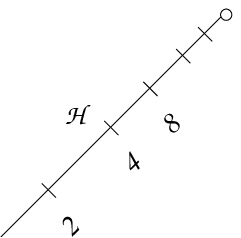}
\]
In each dyadic interval, 
we can select, again as in~\cite{mi:mazi}, a subinterval $A$,
split into two pieces as shown below, each of length
$\sim v^{\frac12}$, 
\[
\includegraphics{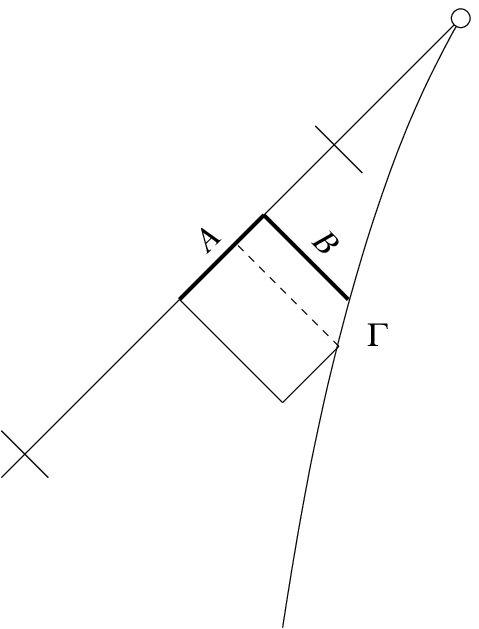}
\]
such that 
\[
\int_A{\theta^2}\le C_1v^{-\left(2q+\frac12\right)}.
\] 
Here $v$ denotes the $v$-value of any point in the dyadic interval.
(Recall that these values are all comparable
to the length of the dyadic interval.)
We note that the assumptions of Propositions of Section 6 of~\cite{mi:mazi} hold. We
thus obtain the bound
\[
\left|\frac{\zeta}{\nu}\right|\le C_2v^{-\left(q+\frac14\right)}
\]
on the interval $B$. It follows
immediately, in view of 
the bounds $c\le r\le R$, and $(\ref{asteri})$
that, 
\[
\int_B{\left(\frac{\zeta}\nu\right)^2(1-\mu)(-\nu)du}\le 
C_3v^{-\left(2q+\frac12\right)}.
\]
Now, by the inequalities $\partial_v\varpi\ge0$, $\partial_u\varpi\le0$,
the fact that $m$ is nonincreasing to the future along $\Gamma$,
and the equation $(\ref{pvqu})$, this implies immediately 
\begin{equation}
\label{evergeia2}
\int_v^\infty\frac{1-\mu}{\lambda}\theta^2(\tilde{U},\bar{v})d\bar{v}
\le C_4v^{-\left(2q+\frac12\right)}
\end{equation}
along the event horizon. By the dyadic nature of the decomposition,
 $(\ref{evergeia2})$
holds for all values of $v$, and not just for the $v$-values of the
future endpoints
of $A$. Finally,
by
$(\ref{alloasteri})$,
this proves immediately
$(\ref{evergeia})$.
Applying again the results of Section 6 of~\cite{mi:mazi}, 
now with \emph{any} subinterval
of the event horizon of length
$\sim v$, in view of $(\ref{evergeia})$,
we 
obtain that the inequality
\begin{equation}
\label{ptwise}
\left|\frac\zeta\nu\right|\le C_5v^{-\left(q+\frac14\right)}
\end{equation}
holds throughout $\mathcal{Q}$, in particular, on
$\Gamma$.
The boundary conditions, namely $\lambda=1$, $R=C$,
$m$ is non-increasing on $\Gamma$,
yield that
\[
|\theta(u,v)|\le\left|\frac\zeta\nu\right|
\]
on $\Gamma$,
and thus by $(\ref{ptwise})$ we have
\[
|\theta|\le C_5v^{-\left(q+\frac14\right)}
\]
on $\Gamma$.
Integrating now the equation 
\[
\partial_u\theta=-\frac\zeta\nu\frac{\lambda}{1-\mu}(1-\mu)r\nu
\]
from $\Gamma$,
in view of the bounds $(\ref{asteri})$, $(\ref{alloasteri})$,
and $(\ref{ptwise})$,
we obtain
\[
|\theta|\le C_6v^{-\left(q+\frac14\right)}
\]
throughout $\mathcal{Q}$. This concludes the proof. $\Box$

\end{document}